\documentclass[showpacs,a msmath,amssymb,aps]{revtex4-1}
\usepackage{graphicx}
\usepackage{subfigure}
\usepackage{tikz}
\definecolor{gold}{rgb}{0.85,0.65,0}
\usepackage{dcolumn}
\usepackage{bm}

\DeclareMathOperator{\Tr}{Tr}

\begin{document}

\title{Charge Induced Fluctuation Forces in Graphitic Nanostructures }

\author{D. Drosdoff$^{1,2}$, Igor V. Bondarev$^{2}$, Allan Widom$^{3}$, Rudolph Podgornik$^{4,5}$, and Lilia M. Woods$^{1}$}
\affiliation{$^{1}$Department of Physics, University of South Florida, Tampa FL 33620}
\affiliation{$^{2}$Department of  Physics, North Carolina Central University, Durham NC 27707}
\affiliation{$^{3}$Department of Physics, Northeastern University, Boston MA 02115}
\affiliation{$^{4}$Department of Theoretical Physics, Jozef Stefan Institute, SI-1000 Ljubljana, Slovenia}
\affiliation{$^{5}$Department of Physics, University of Massachusetts, Amherst, MA 01003}

\begin{abstract}
Charge fluctuations in nano-circuits with capacitor components are shown to give rise to a novel type of long-ranged interaction, which co-exist with the regular Casimir/van der Waals force. The developed theory distinguishes between thermal and quantum mechanical effects, and it is applied to capacitors involving graphene nanostructures. The charge fluctuations mechanism is captured via the capacitance of the system with geometrical and quantum mechanical components.
The dependence on the distance separation, temperature, size, and response properties of the system shows that this type of force can have a comparable and even dominant effect to the Casimir interaction.  Our results strongly indicate that fluctuations induced interactions due to various thermodynamic quantities can have important thermal and quantum mechanical contributions at the micro- and nanoscale. 
\end{abstract}

\maketitle

\section{Introduction}
Fluctuation-induced interactions have widespread applications in materials and micro- and nano-scaled devices. The much studied  Casimir and van der Waals interactions are due to electromagnetic modes fluctuations captured via the dielectric and magnetic response properties of the objects\cite{Casimir48, Lifshitz61}. Such forces are universal, and they are especially prominent at micron distances and below. Fluctuations of many other observables are also possible, which may give rise to different interactions \cite{Drosdoff2006}. In particular, voltage fluctuations  in capacitor systems \cite{Einstein1907} and wires\cite{Johnson1928} have been of much interest, especially for the operation of devices\cite{Homer2013}. Charge fluctuations and the induced forces are of relevance to biological and chemical matter. Specifically, it has been shown that thermal charge fluctuations in ionic solutions can generate an attractive long-ranged dispersive force even between molecules charged with the same sign \cite{Kirkwood1952},\cite{Podgornik1998}.  Fluctuations originating from charge disorder on neutral slabs have been shown to give an additional contribution to the net interacting force while completely masking the typical Casimir-van der Waals interaction\cite{Naji2010}.

The isolation of single graphene layers and synthesis of related nanostructures, such as carbon nanotubes and graphene nanoribbons (GNRs), have brought new directions in electromagnetic fluctuation induced phenomena. For example, the Casimir/van der Waals force involving graphene systems has non-trivial scaling behavior, which can be tuned via temperature, chemical potential and doping modifications\cite{Sernelius2014, Tkatchenko2013, Dobson2006}. Understanding the electromagnetic fluctuations in the context of building a fundamental understanding and making technological designs for such materials can hardly be overestimated.  

Capacitor systems involving graphitic nanostructures have also been studied extensively. The application of a bias on a graphene system above a substrate is of great relevance to many quantum devices. A key factor has been the control of the local graphene electrochemical potential for the device functionality. Capacitance measurements can also be utilized as means to probe the basic electronic properties of graphitic nanostructures and  to learn about their Dirac-like nature. In particular, the 2D  character of graphene is exhibited in the quantum capacitance\cite{Yu2013},\cite{Ilani2006}, a concept describing the penetration of the electrostatic field due to partial screening \cite{Luryi1987}. Despite the presence of charges and voltage bias in graphene based capacitive systems, their fluctuations and subsequent induced interaction effects have never been considered. In this paper, we show that fluctuating charges in a capacitor system are governed by fluctuation-dissipation relations, giving rise to a novel long-ranged  interaction.  The geometrical size of the capacitor is found to play an important role, as it can be used as an effective way to modulate this interaction.

The charge induced dispersive interactions are presented via a general theory utilizing the capacitance concept. We distinguish between thermal and quantum mechanical effects as the characteristic dependences on distance, temperature, and other factors are shown. This Casimir-like force is examined in graphene based capacitors and compared to the typical Casimir interaction.

\begin{figure}[ht]
\includegraphics[scale=0.6]{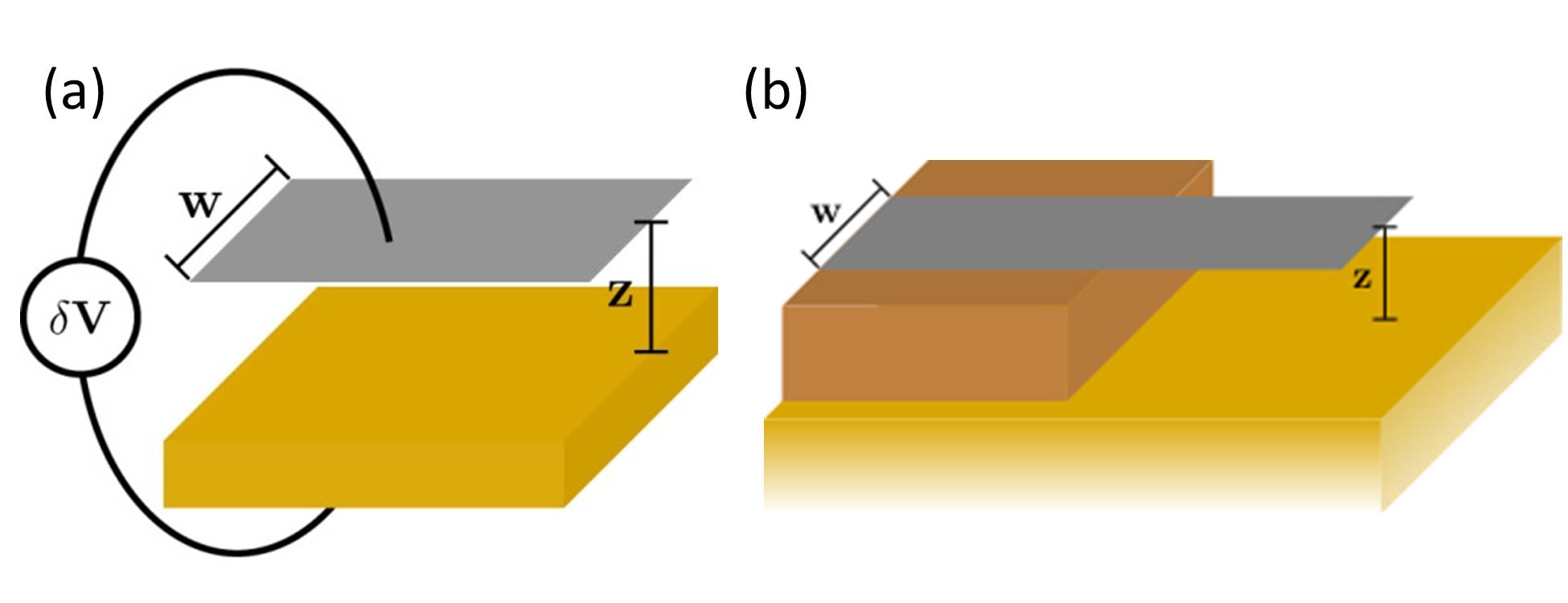}
\caption{(Color online) (a) Ideal system under consideration: A parallel plate capacitor connected by a wire supporting voltage fluctuations $\delta V$. One of the plates can be a graphene ribbon with a width $w$ above a metal substrate separated by a distance $z$. (b) Experimentally realistic setting:  The wire is represented by a metallic connection between the ``bottom'' metal plate and the ``top" graphene nanoribbon.}
\label{fig1}
\end{figure}

\section{Charge Fluctuation Interactions and Capacitance}
The capacitor system under consideration is shown in Fig. 1. It consists of two plates connected by a wire. The voltage fluctuations of the wire induce excess fluctuating charges on the capacitor. As a result, a force originating from these type of fluctuations is induced between the plates. The description of the interaction is closely related to the surface charge density response $\delta\sigma$ due to the external voltage $\delta V$ via the interaction Hamiltonian $\delta H=-\int d^2r'\sigma({\bf r}',t)\delta V({\bf r}',t)$:
  
\begin{equation}
\delta\sigma({\bf r},t)=\int_{-\infty}^{t} dt\int d^2r'C({\bf r},{\bf r}';t-t')\delta V({\bf r}',t'),
\label{cdf2}
\end{equation} 

A key quantity here is the capacitance of the system, which can be written in terms of the commutator of the charge density-density correlations as\cite{Kubo1957}
\begin{equation}
C({\bf r},{\bf r}';t-t')=\frac{i}{\hbar}\langle[\sigma({\bf r},t),\sigma({\bf r}',t')]\rangle,
\label{cdf7}
\end{equation}
where $\langle...\rangle=\Tr(\rho...)$ with $\rho$ being the unperturbed density matrix. 

The density correlation can further be represented as
\begin{equation}
\frac{1}{2}\langle\delta\sigma({\bf r},t)\delta\sigma({\bf r}',t')+ \delta\sigma({\bf r}',t')\delta\sigma({\bf r},t)\rangle=\int\frac{d\omega}{2\pi}\bar{S}({\bf r},{\bf r}';\omega)e^{-i\omega(t-t')},
\label{cdf8}
\end{equation}
and the fluctuation-dissipation theorem is utilized via the structure factor $\bar{S}({\bf r},{\bf r}';\omega)=\hbar\Im m C({\bf r},{\bf r}';\omega)\coth\left(\frac{\hbar\omega}{2k_BT}\right)$. Thus the charge fluctuations are given in terms of Matsubara frequencies $\omega_n=2\pi n k_BT/\hbar$
\begin{equation}
\langle\delta\sigma^2\rangle = k_BT\sum_{n=-\infty}^{\infty}\lim_{{\bf r}\rightarrow {\bf r}'}C({\bf r},{\bf r}';i|\omega_n|)
=k_BT\sum_{n=-\infty}^{\infty}\lim_{{\bf r}\rightarrow {\bf r}'}\int\frac{d^2k}{(2\pi)^2}C({\bf k};i|\omega_n|)e^{i{\bf k}\cdot({\bf r}-{\bf r}')}, 
\label{cdf14}
\end{equation}
\begin{equation}
\langle\delta\sigma^2\rangle \rightarrow \langle\delta q^2\rangle = k_BT\sum_{n=-\infty}^{\infty}C(i|\omega_n|),
\label{cdf15}
\end{equation}
where $\bf k$ is the 2D wave vector. When the voltage fluctuations are distributed uniformly on the plates, the surface charge density fluctuations can be written in terms of the total charge fluctuations according to Eq.(\ref{cdf15}).

Using the electrostatic energy stored in the capacitor  $U=q^2/2C$ , one finds the force per unit area between the capacitor plates as 
\begin{equation}
\bar {f}=-\frac{1}{2A}\frac{\partial C^{-1}}{\partial z}\langle \delta q^2\rangle=-\frac{k_BT}{2A}\frac{\partial C^{-1}}{\partial z}\sum_{n=-\infty}^{\infty}C(i|\omega_n|),
\label{cff1}
\end{equation}
where $A$ is the area of the plate, and $C(\omega)$ is the capacitance of the system that includes the frequency dependent response of the wire.  One should note that the above expression does not take into account the frequency dependent response of the graphene sheet, for example.
Eq.(\ref{cff1}) constitutes a general expression for a charge-induced fluctuation interaction. This Casimir-like effect is determined by the capacitance of the system. To understand the phenomenon further, the theory is applied to graphene systems with emphasis on thermal and quantum mechanical contributions.

The capacitance is obtained by considering the charging of the capacitor, which leads to redistribution of carriers in space in order to minimize the electrostatic energy. Consequently, for classical systems it is determined by the geometry of the system via the geometrical capacitance $C_0$. For thin films, such as graphene and GNRs, the surface charges cannot completely shield the electrostatic field. This results in  raising of the chemical potential to account for the increased density of states necessary for complete shielding \cite{Kopp:2009}. 
The effect is associated with the quantum capacitance $C_Q$, and the total capacitance $C$ is determined by an addition of capacitors in series according to
\begin{equation}
\frac{1}{C}=\frac{1}{C_0}+\frac{1}{AC_Q}.
\label{grc1}
\end{equation}
For the rectangular system in Fig. 1 the geometrical capacitance can be found\cite{Shylau2009} as 
\begin{equation}
C_0=\frac{A}{4}\left[2z\arctan\left(\frac{w}{4z}\right)+\frac{w}{4}\ln\left\{1+\left(\frac{4z}{w}\right)^2\right\}\right]^{-1}.
\label{cgc1}
\end{equation}
One notes that when the  width of the ribbon $w$ is large, the parallel plate capacitance is recovered $\bar{C}_0=A/(4\pi z)$.

The quantum capacitance is associated with the particular material. For graphene it may be obtained via the random phase approximation \cite{Lam2014}, where the particle density is determined from the density matrix $\rho$ and the particle density operator - $n({\bf r})=Tr\left[\delta\rho\delta({\bf r}-{\bf r}_1)\right]$. Using the graphene band structure described with a linear two band model near the characteristic K point, $E_{{\bf k}, s}=s v_0 \hbar k$ with $v_0=10^6$ $m/s$, it is found that
\begin{equation}
C_{Q}({\bf q}, \omega)=-\frac{2e^2}{A}\sum_{{\bf k},s s'}\frac{\left(f_{{\bf k}+{\bf q} s'}-f_{{\bf k} s}\right)|( s' {\bf k}+{\bf q}|{\bf k} s)|^2}{E_{{\bf k}+{\bf q}, s'}-E_{{\bf k}, s}-\hbar\omega},
\label{GC15}
\end{equation}
where $\mu$ is the chemical potential, $s,s'=\pm$, $f_{{\bf k},s}=\frac{1}{e^{(E_{{\bf k},s}-\mu)/k_BT}+1}$, and $( s' {\bf k}+{\bf q}|{\bf k} s)$ is the overlap integral between the wave functions \cite{Brey2007}.  
In the static limit $(\omega, ${\bf k}$)\rightarrow 0$, which is of relevance here, one  finds \cite{Fang2007,Xia2009}
\begin{equation}
C_Q=\frac{ge^2k_BT}{\pi(\hbar v_0)^2}\ln\left[2\cosh\left(\frac{\mu}{2k_BT}\right)\right]=e^2\frac{\partial(n_e-n_h)}{\partial\mu},
\label{GC5}
\end{equation}
where $n_e,\ n_h$ are the electron and hole concentrations respectively.
The above expression reflects the fact that there is only a finite amount of charge available on the graphene sheet, $Q=e(n_e-n_h)$, due to the reduced dimensionality of graphene and its subsequent restriction on the density of states.  Using Eqs.(\ref{grc1},\ref{GC5}), an effective penetration distance may be defined for graphene as $d^*=(\hbar v_0)^2/\left[4ge^2k_BT\ln(2\cosh(\mu/2k_BT))\right]$.  At room temperature with no chemical potential, $d^*\approx 1\ nm$.

The quantum capacitance for GNRs can be obtained in a similar manner. For thin ribbons, however, the energy band structure is quantized due to the vanishing wave functions at the edges \cite{Brey2007}. The effect of the energy quantization in the quantum capacitance can be seen in Fig. 2a for an armchair GNR with $w\approx 12$  $nm$. While the graphene $C_Q$ monotonically increases as a function of $\mu$, the ribbon $C_Q$ experiences peaks \cite{Shylau2009}. This peak structure is much reduced if $T$ is raised to room temperature.

\begin{figure}[ht]
\includegraphics[scale=0.6]{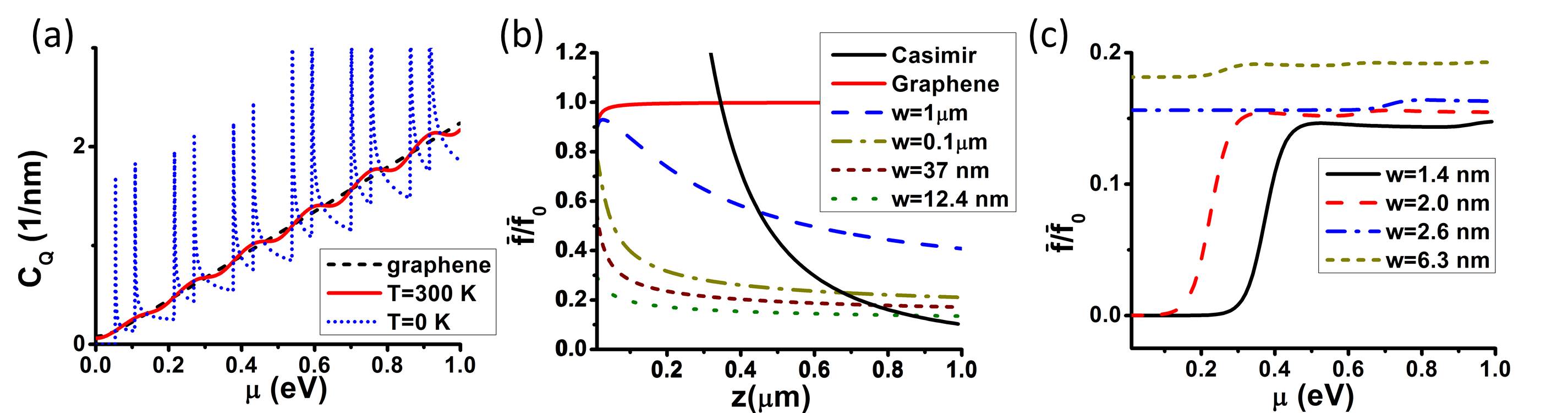}
\caption{(Color online) (a) $C_Q$ as a function of chemical potential for graphene and a GNR with $w=12.4$ $nm$. (b) Charge fluctuations thermal force normalized by $\bar {f}_{0}=-k_BT/(2Az)$. The area of all ribbon structures is $1\  \mu m^2$ and $\mu=0$. The  graphene/metal Casimir force normalized by $\bar {f}_0$ at room temperature is also shown. The metal plasma frequency is taken as $\omega_p=9$ $eV$. (c) The normalized charge fluctuations thermal force vs chemical potential for different GNRs at $T=300$ $K$ and separation $0.1\ \mu m$.}
\label{fig2}
\end{figure}

\section{Results and Discussion}

The charge fluctuations force is determined by the total capacitance, which contains the frequency dependent quantum capacitance according to Eq. (\ref{cff1}). We distinguish between thermal and quantum mechanical contribution in a similar manner as in the case of the typical Casimir force. The thermal $n=0$ Matsubara frequency is considered first. The force per unit area for the capacitor with a thicker GNR, in which the discrete band structure is not apparent,  (Fig. 1) is found analytically

\begin{equation}
{\bar f}_T=-\frac{4k_BT}{A^2}{C_0(z)\cdot\arctan\left(\frac{w}{4z}\right)}\frac{\frac{ge^2k_BT}{\pi(\hbar v_0)^2}\ln\left[2\cosh\left(\frac{\mu}{2k_BT}\right)\right]}{C_0(z)/A +\frac{ge^2k_BT}{\pi(\hbar v_0)^2}\ln\left[2\cosh\left(\frac{\mu}{2k_BT}\right)\right]}.
\label{force2}
\end{equation}
For a graphene parallel plate capacitor, the geometrical capacitance is replaced by $\bar {C}_0=A/4\pi z$ and $w\rightarrow\infty$ in ${\bar f}_T$.

The characteristic behavior of the thermal charge fluctuations interactions can now be determined using Eq.(\ref{force2}). Fig. \ref{fig2}b shows how the thermal force depends on the separation for both graphene and GNRs with different widths. 
For separations $z>>d^*$, one finds that $C_Q>>C_0$, and the thermal force is determined mainly by the geometrical capacitance. In the case of graphene, the force reduces to ${\bar f}_{0}=-\frac{k_BT}{2A}\frac{1}{z}$. For separations $z<<d^*$, $\bar{f}_T$ is determined mainly by the quantum capacitance.  For thinner ribbons, the magnitude of the force is reduced,  as seen in Fig. \ref{fig2}b. While the effect of $C_Q$ for graphene is relatively small over a large part of the distance range $z$, its role for thinner ribbons can be much enhanced (Fig. 2c). This feature, which reduces the magnitude of $f_T$, is attributed to the quantized band structure of the thin ribbons. 

It is important to compare how the typical Casimir force differs from the charge fluctuations induced interaction.  The Casimir graphene/metallic substrate  interaction has been calculated by the Lifshitz approach \cite{Sernelius2014}. In Fig. 2b we show results for an infinite graphene sheet, described by the Dirac model, and a typical metal with a plasma frequency of $\omega_p=9\ eV$.   Due to its $1/z$ dependence, however, $f_T$ is more important for larger $z$.  It should be noted that the region where the size of the Casimir force is comparable to the charge fluctuation force is in that separation length where the thermal Casimir force is dominant.  This reduction in the distance scale where thermal effects take place, as compared to other systems, is due\cite{Gomez2009} to the reduced dimensionality of graphene.

Fig. \ref{fig2}c shows how the thermal charge fluctuations force evolves as a function of $\mu$ for GNRs with different widths. Due to the quantized electronic structure, $\mu$ affects the force much more for thinner ribbons. For thicker ribbons, however, the interaction is hardly changed upon $\mu$. 

\begin{figure}[ht]
\includegraphics[scale=0.6]{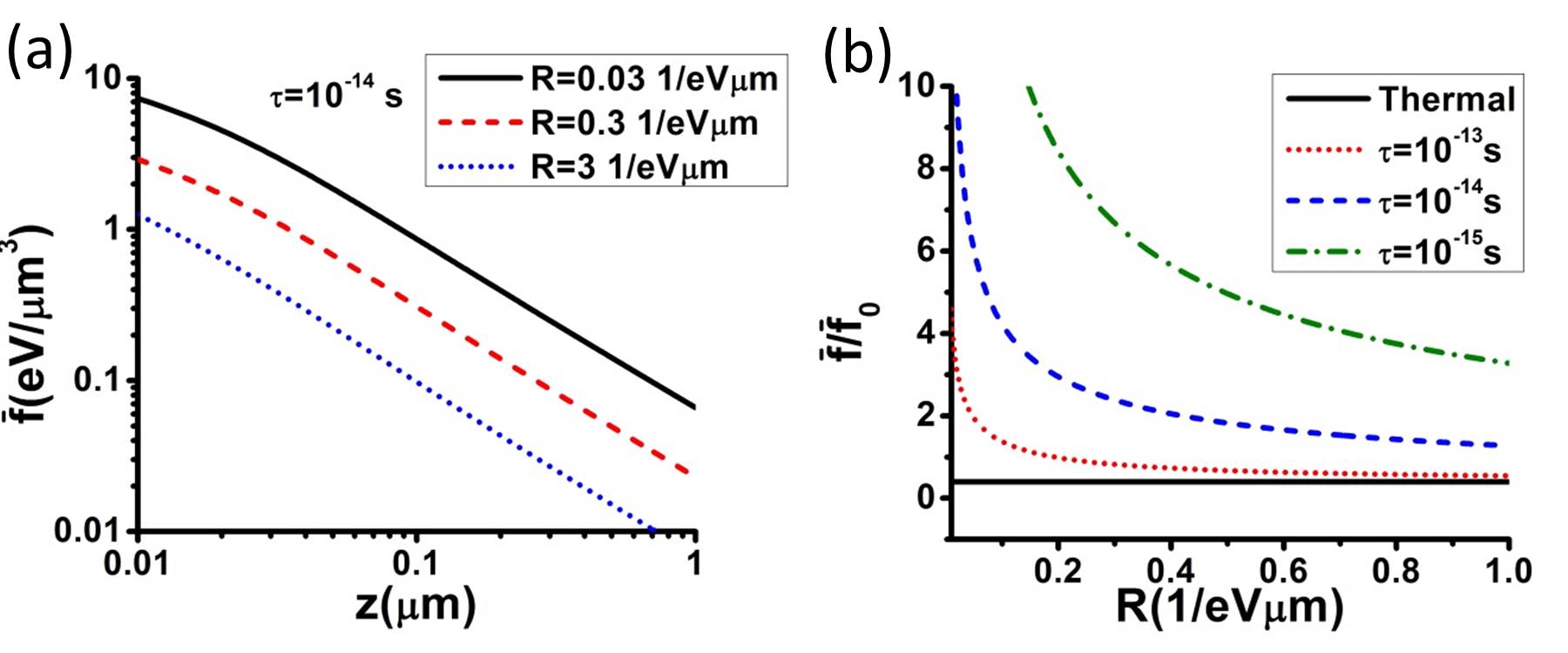}
\caption{(Color online) (a) Charge fluctuations force vs. distance between ribbon of width 0.1 $\mu m$ and area 1 $\mu m^2$ and a metal plate at room temperature, where the $C^t(\omega)$ is used. Typical values for $\tau$ and $R$ are taken. (b) The normalized by $\bar {f}_{0}$ charge fluctuations force as a function of the wire resistance for different relaxation times and a plate separation of $0.1\ \mu m$. }
\label{fig3}
\end{figure}

The quantum mechanical contribution to the charge fluctuations force is closely related to the frequency dependent total capacitance.One must account for the particular mechanism for charging the capacitor, which happens via the connecting wire, as shown in Fig. 1. Taking a Drude model for the wire conductivity leads to a frequency dependent total capacitance according to 
\begin{equation}
\frac{1}{C^t (\omega)}=\frac{1}{C_0}+\frac{1}{AC_Q}-i\omega R-\omega^2 R\tau.
\label{qf1}
\end{equation}
where $R$ is the resistance of the wire and $\tau$ is the relaxation time \cite{Johnson1928, Jeon2009}. Using Eqs. (\ref{cff1},\ref{qf1}), the charge fluctuations force can now be calculated by taking into account all Matsubara frequencies. Results for the distance and $R$ dependence of the force are shown in Fig. \ref{fig3}. It is found that the role of $R$ is more pronounced for smaller $z$ as the effect is stronger for smaller $R$ values, according to  Fig. \ref{fig3}a. Similar conclusion is obtained by considering the dependence of $\bar {f}$ vs $R$, shown in Fig. \ref{fig3}b, which indicates that larger resistance (for given $\tau/C_0$) leads to thermal fluctuations dominating the interaction.

The quantum mechanical effects related to the mechanism of charging the capacitor can be analyzed further by considering the situation when $RC_0$ is small compared to $\tau$ and $C_Q$ can be neglected.  The total capacitance in this case becomes $C^t(\omega)\approx\frac{C_0}{1-\omega^2RC_0\tau}$. From Eqs. (\ref{cff1}), the force is found as
\begin{equation}
\bar {f}=-\frac{\hbar}{4Az\sqrt{RC_0\tau}}\coth\left(\frac{\hbar}{2k_BT\sqrt{RC_0\tau}}\right).
\label{TPP7}
\end{equation}
In the classical limit of large $T$, one recovers $\bar {f}_{0}=-\frac{k_BT}{2Az}$. In the quantum mechanical limit, where $\hbar/\sqrt{RC_0\tau}\gg k_BT$, we obtain
\begin{equation}
\bar {f}=-\frac{\hbar}{2A\sqrt{z}}\sqrt{\frac{\pi}{AR\tau}}.
\label{TPP8}
\end{equation}
Eqs. (\ref{TPP7}, \ref{TPP8}) are instrumental in better understanding the $\bar {f}$ dependence upon $z$ in Fig. 3. As the distance is increased, the thermal effects become more important according to the $1/z$ asymptotic behavior. However, for smaller separations, the quantum mechanical contributions become more relevant according to the $1/\sqrt{z}$ asymptotic behavior. In all cases, these effects are more pronounced for smaller $R$, otherwise, the charge induced interaction is determined by the thermal fluctuations.

Casimir and van der Waals forces are universal and of great importance in nature, especially when the scale goes down to micrometer distances and below. This type of interaction, induced by dipolar fluctuations, can be essential in determining the behavior of nanostructures and devices.  Yet there are many other fluctuation induced forces that become important at such small scales. Here we consider charge induced fluctuation forces in capacitive systems. The developed theory utilizes  static and frequency dependent capacitance, and it is applied in graphene based structures.  The total capacitance contains both a geometrical and a collective electronic component. Just like in the Casimir/van der Waals forces, we distinguish between thermal and quantum mechanical effects. The thermal fluctuations are pronounced at larger separations where the geometrical capacitance dominates, while the quantum fluctuations are relevant for smaller separations.

The origin of this induced interaction comes from fluctuating charges, which is a key difference when compared to the Casimir/van der Waals interaction that originates from fluctuating polarizations. This is important for understanding the $1/z$ asymptotic distance dependence as compared to the $1/z^3$ dependence of the thermal Casimir force. Similarly, the quantum mechanical regime of $1/\sqrt{z}$ is of much longer range as compared to the metal/graphene quantum mechanical asymptotics \cite{Sernelius2014}, \cite{Dobson2006}. It is also important to note that the geometrical capacitance takes into account the geometry of the system  in a rather straight forward manner. This is unlike the Casimir force, where non-trivial geometry is usually difficult to describe theoretically. This is reflected in the fact that capacitance can be found for objects with non-trivial boundary conditions via electrostatic methods, while electromagnetic boundary conditions are not easily solved beyond highly symmetric extensions. 

Another interesting feature is the role of the size of the system. The charge fluctuations force is explicitly dependent on the area of the interacting objects, which is not the case for the Casimir/vdW force.The charge fluctuations interaction is $\sim 1/A$ for the thermal and $\sim 1/A^{3/2}$ for the quantum mechanical regimes. Balancing the longer ranged asymptotics and the area dependence  can be used to enhance the charge fluctuations force over the typical Casimir interaction. Systems, such as graphene ribbons, which have smaller area as compared to large graphene flakes, are especially good candidates to observe this effect.  

In summary, we argue that charge fluctuation forces are always present in capacitor-like systems. They must be considered in conjunction with the standard Casimir interaction as the charge fluctuation force can be comparable or even bigger in magnitude. Charge fluctuation forces enable further probing of thermal and quantum mechanical effects 
due to fluctuation induced phenomena in nanostructured materials such as graphene and GNRs.  We finally point out that it is worthwhile to pursue further the effects of fluctuation forces due to not just electromagnetic fluctuations but also due to other thermodynamic parameters, for they will necessarily have an important effect in many nanostructures and devices.

\section{Acknowledgements}

We acknowledge financial support from the Department of Energy under Contract No. DE-FG02-06ER46297. I.V.B. acknowledges support from the National Science Foundation under Contract No. ECCS-1306871.

\end{document}